\documentclass[preprint,aps,12pt,showpacs,nofootinbib,tightenlines]{revtex4}
\usepackage{mathrsfs}
\usepackage{amsmath}
\usepackage{amssymb}
\usepackage{epsfig}
\usepackage{graphicx}
\newcommand{\psl}{ P \hspace{-2.4truemm}/ }

\newcommand{\esl}{ \epsilon \hspace{-1.5truemm}/ }

\def\be{\begin{eqnarray}}
\def\en{\end{eqnarray}}
\def\non{\nonumber\\}

\def\prd{{Phys. Rev. D}~}
\def\prl{{ Phys. Rev. Lett.}~}
\def\plb{{ Phys. Lett. B}~}

\begin{document}
%%--------------------------------------------
\title{Study of the $K_1(1270)-K_1(1400)$ mixing in the decays $B\to J/\Psi K_1(1270), J/\Psi K_1(1400)$}
\author{Zhi-Qing Zhang$^1$, Hongxia Guo\footnote{Corresponding author: Hongxia Guo, e-mail: guohongxia@zzu.edu.cn.}$^2$, Si-Yang Wang$^1$}
\affiliation{\it \small $^1$  Department of Physics, Henan
University of
Technology,\\ \small \it Zhengzhou, Henan 450052, P. R. China; \\
\small $^2$ \it School of Mathematics and Statistics, Zhengzhou University,\\
\small \it Zhengzhou, Henan 450001, P. R. China } %%
\date{\today}
\begin{abstract}
We study the B meson decays $B\to J/\Psi K_1(1270,1400)$ in the pQCD
approach beyond the leading order. With the vertex corrections and
the NLO Wilson coefficients included, the branching ratios of the
considered decays are predicted as $Br(B^+\to J/\Psi
K_1(1270)^+)=1.76^{+0.65}_{-0.69}\times10^{-3}, Br(B^+\to J/\Psi
K_1(1400)^+)=6.47^{+2.50}_{-2.34}\times10^{-4}$, and $Br(B^0\to
J/\Psi K_1(1270)^0)=(1.63^{+0.60}_{-0.64})\times10^{-3}$ with the
mixing angle $\theta_{K_1}=33^\circ$, which can agree well with the
data or the present experimental upper limit within errors. So we
support the opinion that $\theta_{K_1}\sim33^\circ$ is much more
favored than $58^{\circ}$. Furthermore, we also give the predictions
of the polarization fractions, the direct CP violations, the
relative phase angles for the considered decays with the mixing
angle $\theta_{K_1}=33^\circ$ and $58^\circ$, respectively. The
direct CP violations of the two charged decays $B^+\to J/\Psi
K_1(1270,1400)^+$ are very small $(10^{-4}\sim10^{-5})$, because the
weak phase is very tiny. In order to check the dependence of the
results on the nonperturbative input parameters, we also calculate
them by using the harmonic-oscillator type wave functions for the
$J/\Psi$ meson. These results can be tested at the running LHCb and
forthcoming Super-B experiments.
\end{abstract}

 \vspace{1cm}

\maketitle

%=======================================================================
%                     Introduction
%=======================================================================

\section{Introduction}\label{intro}
$B$ meson exclusive decays into charmonia have been received a lot
of attentions for many years. They are regarded as the golden
channels in researching CP violation and exploring new physics. At
the same time, they play the important roles in testing the
unitarity of the Cabibbo-Kobayashi-Maskawa (CKM) triangle. Moreover,
these decays are ideal modes to test the different factorization
approaches. Compared with other factorization approaches, such as
the naive factorization assumption (FA) \cite{ali}, the QCD-improved
factorization (QCDF) \cite{beneke}, the perturbtive QCD (pQCD)
approach \cite{lihn1} has the unique advantage in solving the B
meson charmed decays \cite{chen,zhang}. The Sudakov factor induced
by the $k_T$ resummation \cite{lihn} can eliminate the double
logarithmic divergences.  The jet function induced by the threshold
resummation \cite{lihn2} can smear the end-point singularities.
Without the divergences, one can evaluate all possible Feynman
diagrams correctly, including the nonfactorizable emission diagrams
and annihilation type diagrams. But it is difficult to calculate
these two kinds of contributions by using other factorization
approaches.

Some of the decays $B\to J/\Psi K_1(1270), J/\Psi K_1(1400)$ have
been measured by Belle\cite{abe}, \be
Br(B^+\to J/\Psi K^+_1(1270))&=&(1.80\pm0.34\pm0.39)\times10^{-3},\\
Br(B^+\to J/\Psi K^+_1(1400))&<&5.4\times10^{-4},\\
Br(B^0\to J/\Psi K^0_1(1270))&=&(1.30\pm0.34\pm0.31)\times10^{-3},
\en where the first uncertainties are statistical and the second are
systematic.

As we have known, the physical mass eigenstates $K_1(1270)$ and
$K_1(1400)$ are the mixing by the flavor eigenstates $K_{1A}$ and
$K_{1B}$ through the following formula \be \left(\begin{matrix}
|K_1(1270)\rangle\\
|K_1(1400)\rangle
\end{matrix}\right)=\left(\begin{matrix}\sin\theta_{K_1} & \cos\theta_{K_1}\\
                           \cos\theta_{K_1} & -\sin\theta_{K_1}\end{matrix}\right)
                           \left(\begin{matrix}|K_{1A}\rangle \\ |K_{1B}\rangle\end{matrix}\right).
\label{mixing} \en Usually we combine $K_{1A}$ with $a_1(1260),
f_1(1285), f_1(1420)$ to form the nonet $J^{PC}=1^{++}$, while
combine $K_{1B}$ with $b_1(1235), h_1(1170), h_1(1380)$ to comprise
the other nonet $J^{PC}=1^{+-}$. These two nonet mesons can also be
denoted as $^3P_1$ and $^1P_1$ in terms of the spectrosocpic
notation $^{2S+1}L_{J}$. Various phenomenological studies indicate
that the mixing angle $\theta_{K_1}$ is around either $33^\circ$ or
$58^\circ$ \cite{suzuki,bura,cheng,yang0,god,hata,tay,div}.

In view of the above situation, the motivations are in order: (a)
Proving whether the pQCD approach can be used in our considered
decays by comparing with the data. Several earlier works on $B$
decays into charmonia \cite{chen,cdlu1,liu0} show that this approach
can give the results in agreement with data, which encourage our
attempt. (b)Exploring the inner structure of the axial vector mesons
$K_1(1270, 1400)$, in other words, detecting which mixing angle
shown in Eq.(4) is favored. (c) Studying of CP violation even new
physics in these decays containing the charmonium state. Besides the
full leading-order (LO) contributions, the next-to-leading-order
(NLO) contributions are also included, which are mainly from the NLO
Wilson coefficients and the vertex corrections to the hard kernel.
Certainly, other NLO contributions, such as the quark loops and the
magnetic penguin corrections, are also available in the
literature\cite{li,xiao}, while they will not contribute to these
considered decays.

We review the LO order predictions for the decays $B\to J/\Psi
K_1(1270),J/\Psi K_1(1400)$ including those for the main NLO
contributions in Section II. We perform the numerical study in
Section III, where the theoretical uncertainties are also
considered. Section IV is the conclusion.
\section{the Leading-Order Predictions and the main next-to-leading order corrections}
The weak effective Hamiltonian $H_{eff}$ for the decays $B\to J/\Psi
K_1(1270, 1400)$ can be written as: \be
\emph{H}_{eff}=\frac{G_F}{\sqrt2}\left[V^*_{cb}V_{cs}(C_1(\mu)O^c_1(\mu)+C_2(\mu)O^c_2(\mu))-
V^{*}_{tb}V_{ts}\sum^{10}_{i=3}C_i(\mu)O_i(\mu)\right], \en where
$C_i(\mu)$ are Wilson coefficients at the renormalization scale
$\mu$, $V$ represents for the Cabibbo-Kobayashi-Maskawa (CKM) matrix
element, and the four fermion operators $O_i$ are given as: \be
O^c_1&=&(\bar s_{i}c_{j})_{V-A}(\bar c_{j}b_{i})_{V-A},\;\;\;
O^c_2=(\bar s_{i}c_{i})_{V-A}(\bar
c_{j}b_{j})_{V-A},\\
O_3&=&(\bar s_{i}b_{i})_{V-A}(\bar q_{j}q_{j})_{V-A},\;\;\;
O_4=(\bar s_{i}b_{j})_{V-A}(\bar
q_{j}q_{i})_{V-A},\\
O_5&=&(\bar s_{i}b_{i})_{V-A}(\bar q_{j}q_{j})_{V+A},\;\;\;
O_6=(\bar s_{i}b_{j})_{V-A}(\bar
q_{j}q_{i})_{V+A},\\
O_7&=&\frac{3}{2}(\bar s_{i}b_{i})_{V-A}\sum_{q}e_{q}(\bar
q_{j}q_{j})_{V+A},\;\;\; O_8=\frac{3}{2}(\bar
s_{i}b_{j})_{V-A}\sum_{q}e_{q}(\bar
q_{j}q_{i})_{V+A},\\
O_9&=&\frac{3}{2}(\bar s_{i}b_{i})_{V-A}\sum_{q}e_{q}(\bar
q_{j}q_{j})_{V-A},\;\;\; O_{10}=\frac{3}{2}(\bar
s_{i}b_{j})_{V-A}\sum_{q}e_{q}(\bar q_{j}q_{i})_{V-A}, \en with
$i,j$ being the color indices.
\begin{figure}[t]
\vspace{-4cm} \centerline{\epsfxsize=18 cm \epsffile{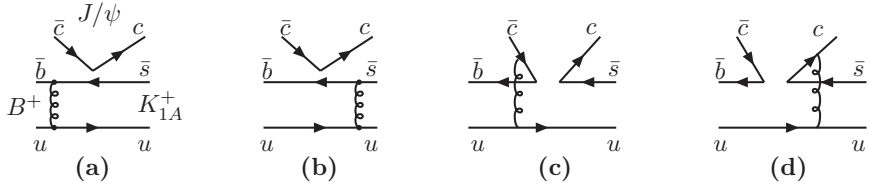}}
\vspace{-18.8cm} \caption{ Feynman Diagrams contributing to the
decay $B^+\to J/\Psi K^+_{1A}$ at the leading order. The hard gluon
connecting the four quark operator and the spectator quark is
necessary to ensure the pQCD applicability. They are the same with
those for $B^+\to J/\Psi K^+_{1B}$. If replacing the spectator $u$
quark with $d$ quark, we will obtain the Feynman Diagrams for the
decays $B^0\to J/\Psi K^0_{1A}, J/\Psi K^0_{1B}$. }
 \label{Figure1}
\end{figure}

It is convenient to do the calculation in the rest frame of $B$
meson because of the heavy $b$ quark. Throughout this paper, we take
the light-cone coordinate $(P^+,P^-,\textbf{P}_T)$ to describe the meson's
momenta with $P^{\pm}=(p_0\pm p_3)/\sqrt2$ and
$\textbf{P}_T=(p_1,p_2)$. Then the momenta of mesons $B, J/\Psi$ and
$K_1$ can be written as \be
P_1=\frac{m_B}{\sqrt2}(1,1,\textbf{0}_T),
\;\;\;\;\;P_2=\frac{m_B}{\sqrt2}(1-r^2_3,r^2_2,\textbf{0}_T),
\;\;\;\;\; P_3=\frac{m_B}{\sqrt2}(r^2_3,1-r^2_2,\textbf{0}_T), \en
respectively. The mass ratios $r_2=m_{J/\Psi}/m_B, r_3=m_{K_1}/m_B$.
In the numerical calculation, the terms proportional to $r^2_3$ are
neglected, as $r^2_3\sim 0.06$ is numerically small. Putting the
(light) quark momenta in $B, J/\Psi, K_1$ mesons as $k_1, k_2, k_3$,
respectively, we have \be k_1=(x_1P^+_1,0,\textbf{k}_{1T}),
\;\;\;\;k_2=x_2P_2+(0,0,\textbf{k}_{2T}),
\;\;\;\;k_3=x_3P_3+(0,0,\textbf{k}_{3T}). \en

There are three kinds of polarizations of a vector or an
axial-vector meson, namely longitudinal (L), normal (N) and
transverse (T). So the amplitudes for the decay mode $B(P_1)\to
V_2(P_2,\epsilon^*_{2\mu}) +A_3(P_3,\epsilon^*_{3\nu})$ are
characterized by those polarization states, which can be decomposed
as follows: \be
A^{(\sigma)}&=&\epsilon^*_{2\mu}(\sigma)\epsilon^*_{3\nu}(\sigma)\left[ag^{\mu\nu}+\frac{b}{M_2
M_3}P^\mu_1P^\nu_1
+i\frac{c}{M_2M_3}\epsilon^{\mu\nu\alpha\beta}P_{2\alpha}P_{3\beta}\right]\non
&&
=\mathcal{M}^L+\mathcal{M}^N\epsilon^*_2(\sigma=T)\cdot\epsilon^*_3(\sigma=T)
+i\frac{\mathcal{M
}^T}{M^2_B}\epsilon^{\alpha\beta\gamma\rho}\epsilon^*_{2\alpha}(\sigma)
\epsilon^*_{3\beta}(\sigma)P_{2\gamma}P_{3\rho}, \en where
$M_{2(3)}$ is the mass of the vector (axial-vector) meson
$V_2(A_3)$. The definitions of the amplitudes
$\mathcal{M}^j(j=L,N,T)$ in terms of the Lorentz-invariant
amplitudes $a, b$ and $c$ are given as: \be
\mathcal{M}^{L}&=&a \epsilon^*_{2}(L)\cdot\epsilon^*_{3}(L)+\frac{b}{M_2M_3}\epsilon^*_{2}(L)\cdot P_3 \epsilon^*_{3}(L)\cdot P_2,\\
\mathcal{M}^N  &=&a, \;\;\;\; \mathcal{M}^T=\frac{c}{r_2r_3}. \en
 It is noticed that the
subscript $K_1$ refers to the flavor eigenstate $K_{1A}$ or
$K_{1B}$. At the leading order, the relevant contributions are only
from the factorizable and non-factorizable emission diagrams, as
shown in Fig.1. We take the decay $B^+\to J/\Psi K_{1A(B)}^+$ as an
example. The emission particle is the vector meson $J/\Psi$, and the
amplitude for the factorizable emission diagrams Fig.1(a) and
Fig.1(b) from the longitudinal polarization can be written as: \be
\mathcal{F}^{L}_{J/\Psi K_1}&=&8\pi C_F m^4_Bf_{J/\Psi}\int_0^1 dx_1
dx_3 \int_0^\infty b_1 db_1 b_3 db_3 \phi_B(x_1,b_1)(r^2_2-1)\non &&
\left\{\left[(1+(1-r^2_2)x_3)\phi_{K_1}(x_3)
+r_{3}(1-2x_3)(\phi^s_{K_1}(x_3)+\phi^t_{K_1}(x_3))\right]\right.
\non &&\left.
\times\alpha_s(t_a)E_e(t_a)h_e(x_1,x_3,b_1,b_3)S_t(x_3)\right.\non
&&\left.
+\alpha_s(t_b)E_e(t_b)h_e(x_3,x_1,b_3,b_1)S_t(x_1)2r_{3}\phi^s_{K_1}(x_3)
\right\}, \en where the color factor $C_F=4/3$. $\phi_{K_1}$ and
$\phi^{t,s}_{K_1}$ are the twist-2 and twist-3 distribution
amplitudes for the axial-vector meson $K_{1A}$ or $K_{1B}$, which
can be found in Appendix A. The evolution factors evolving the
Sudakov factor, the hard function $h_{e}$ and the jet function
$S_t(x)$ are given in Appendix B. Similarly, the normal and
transverse polarization amplitudes are displayed as \be
\mathcal{F}^{N}_{J/\Psi K_1}&=&8\pi C_F m^4_Bf_{J/\Psi}\int_0^1 dx_1
dx_3 \int_0^\infty b_1 db_1 b_3 db_3 \phi_B(x_1,b_1)r_{2}\non &&
\left\{\left[r_{3}((r^2_2-1)x_3((r^2_2-1)\phi^a_A(x_3)+\phi^v_A(x_3))-2\phi^v_A(x_3))
\right.\right. \non &&\left.\left. +(r^2_2-1)\phi^a_T(x_3) \right]
\alpha_s(t_a)E_e(t_a)h_e(x_1,x_3,b_1,b_3)S_t(x_3)\right.\non
&&\left.
-r_{3}\left[(1-r^2_{2})\phi^v_A(x_3)+(r^2_2-1)^2\phi^a_A(x_3)\right]\alpha_s(t_b)E_e(t_b)\right.\non
&&\left. \times h_e(x_3,x_1,b_3,b_1)S_t(x_1)
\right\},\en
\be
\mathcal{F}^{T}_{J/\Psi K_1}&=&-16\pi C_F m^4_Bf_{J/\Psi}\int_0^1
dx_1 dx_3 \int_0^\infty b_1 db_1 b_3 db_3 \phi_B(x_1,b_1)r_2 \non &&
\left\{\left[r_{3}((r^2_2-1)x_3+2)\phi^a_A(x_3)-r_{3}x_3\phi^v_A(x_3)+\phi^T_A(x_3)\right]\right.\non
&& \left.
\alpha_s(t_a)E_e(t_a)h_e(x_1,x_3,b_1,b_3)S_t(x_3)\right.\non
&& \left.
+r_{3}\left[(1-r^2_{2})\phi^a_A(x_3)+\phi^v_A(x_3)\right]\alpha_s(t_b)E_e(t_b)\right.\non
&&\left. \times h_e(x_3,x_1,b_3,b_1)S_t(x_1) \right\}. \en

The longitudinal polarization amplitude for the non-factorizable
spectator diagrams Fig.1(c) and Fig.1(d) is given as:
\be
\mathcal{M}^{L}_{J/\Psi K_1}&=&\frac{32}{\sqrt6}\pi C_F m^4_B
\int_0^1 dx_1 dx_2 dx_3 \int_0^\infty b_1 db_1 b_2 db_2
\phi_B(x_1,b_1)(r^2_{2}-1)\non
&&\times(2r_{3}\phi_{K_1}^t(x_3)-\phi_{K_1}(x_3))\left[2r_cr_{2}\psi^t(x_2)+(r^2_2(x_3-2x_2)-x_3)\psi^L(x_2)\right]
\non && \times \alpha_s(t_d)E_{en}(t_d)h_d(x_1,x_2,x_3,b_1,b_2),
\label{cd1} \en where the twist-2 and twist-3 distribution
amplitudes $\psi^{L,t}(x_2)$ for the $J/\Psi$ meson (Type I) can be
found in Appendix A. The other two polarization amplitudes are
written as: \be \mathcal{M}^{N}_{J/\Psi K_1}&=&\frac{64}{\sqrt6}\pi
C_F m^4_B \int_0^1 dx_1 dx_2 dx_3 \int_0^\infty b_1 db_1 b_2 db_2
\phi_B(x_1,b_1)\non
&&\times\left\{r_2\psi^v(x_2)\left[r_{3}(x_2(1+r_2^2)+x_3(1-r_2^2))
\phi_{K_1}^v(x_3)-x_2(1-r_2^2)\phi^T_{K_1}(x_3)\right]\right.\non
&&\left.
+r_c\psi^T(x_2)\left[(1-r_2^2)\phi^T_{K_1}(x_3)-r_{3}(1+r_2^2)\phi_{K_1}^v(x_3)\right]
\right\}\non && \alpha_s(t_d)E_{en}(t_d)h_d(x_1,x_2,x_3,b_1,b_2),\\
\mathcal{M}^{T}_{J/\Psi K_1}&=&\frac{128}{\sqrt6}\pi C_F m^4_B
\int_0^1 dx_1 dx_2 dx_3 \int_0^\infty b_1 db_1 b_2 db_2
\phi_B(x_1,b_1)\non
&&\times\left\{r_2\psi^v(x_2)\left[r_{3}(x_2(1+r_2^2)+x_3(1-r_2^2))
\phi_{K_1}^a(x_3)-x_2\phi^T_{K_1}(x_3)\right]\right.\non &&\left.
+r_c\psi^T(x_2)\left[\phi^T_{K_1}(x_3)-r_{3}(1+r_2^2)\phi_{K_1}^a(x_3)\right]
\right\}\non && \alpha_s(t_d)E_{en}(t_d)h_d(x_1,x_2,x_3,b_1,b_2).
\en

By combining these amplitudes from the different Feynman diagrams
and Eq.(\ref{mixing}), one can get the total decay amplitude for the
decay $B^+\to J/\Psi K_{1}(1270)^+$: \be \mathcal{M}^{j}(B^+\to
J/\Psi K_{1}(1270)^+)&=&\mathcal{M}^{j}(B^+\to J/\Psi
K^+_{1A})\sin\theta_{K_1}+\mathcal{M}^{j}(B^+\to J/\Psi
K^+_{1B})\cos\theta_{K_1}\non &=&(\mathcal{F}^{j}_{J/\Psi
K_{1A}}\sin\theta_{K_1}+\mathcal{F}^{j}_{J/\Psi
K_{1B}}\cos\theta_{K_1})\non && \times
\left[V^*_{cb}V_{cs}a_2-V^*_{tb}V_{ts}(a_3+a_5+a_7+a_9)\right]\non
&& +(\mathcal{M}^{j}_{J/\Psi
K_{1A}}\sin\theta_{K_1}+\mathcal{M}^{j}_{J/\Psi
K_{1B}}\cos\theta_{K_1}) \non &&
\times\left[V^*_{cb}V_{cs}C_2-V^*_{tb}V_{ts}(C_4-C_6-C_8+C_{10})\right],
\label{totalam} \en where $\mathcal{M}^{j}$ and $\mathcal{F}^{j}$
($j=L,N,T$) refer to the different helicity amplitudes. The
combinations of the Wilson coefficients $a_2=C_1+C_2/3,
a_i=C_i+C_{i+1}/3$ with $i=3,5,7,9$. As for the decays $B^+\to
J/\Psi K_{1}(1400)^+$, the total amplitude can be obtained by
replacing $\sin\theta_{K_1}$ and $\cos\theta_{K_1}$ with
$\cos\theta_{K_1}$ and $-\sin\theta_{K_1}$ in Eq.(\ref{totalam}),
respectively.

Here only the vertex corrections need to be considered in the NLO
calculations for the decays $B^+\to J/\Psi K^+_{1A,B}$. Since the
vertex corrections can reduce the dependence of the Wilson
coefficients on the renormalization scale $\mu$, they usually play
the important roles in the NLO analysis. It is well known that the
nonfactorizable amplitude contributions are small \cite{lihn1}, we
concentrate only on the vertex corrections to the factorizable
amplitudes, as shown in Fig.2. Furthermore, the infrared divergences
from the soft and the collinear gluons in these Feynman diagrams can
be canceled each other. That is to say, these corrections are free
from the end-point singularity in the collinear factorization
theorem, so we can quote the QCDF expressions for the vertex
corrections: their effects can be combined into the Wilson
coefficients, \be
a^h_2&\rightarrow& a_2+\frac{\alpha_s C_F}{4\pi N_c}C_2(-18+12\ln\frac{m_b}{\mu}+f^h_I),\\
a^h_i&\rightarrow& a_{i}+\frac{\alpha_s C_F}{4\pi N_c}C_{i+1}(-18+12\ln\frac{m_b}{\mu}+f^h_I), (i=3,9),\\
a^h_i&\rightarrow& a_{i}+\frac{\alpha_s C_F}{4\pi
N_c}C_{i+1}(6-12\ln\frac{m_b}{\mu}-f^h_I), (i=5,6), \en with the
function $f^h_I(h=0,\pm)$ defined as: \be f^0_I=f_I+g_I(1-r^2),
\;\;\; f^{\pm}=f_I. \en As for the expressions of $f_I$ and $g_I$
are given in Appendix C. Certainly, the NLO Wilson coefficients will
be used in the NLO calculations.
\begin{figure}[]
\vspace{-4cm} \centerline{\epsfxsize=18 cm \epsffile{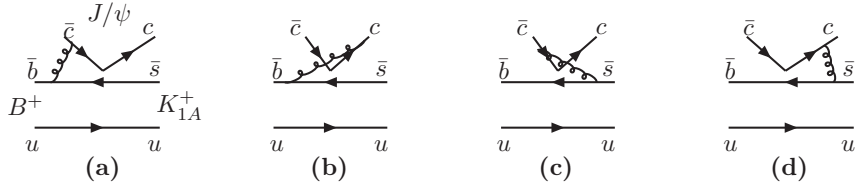}}
\vspace{-18.8cm} \caption{ NLO vertex corrections to the
factorizable emission diagrams Fig.1(a) and (b) for the decay
$B^+\to J/\Psi K^+_{1A}$. Here the hard gluon is not shown for
simplicity.
   It
is the same with those for the decay $B^+\to J/\Psi K^+_{1B}$.}
 \label{Figure2}
\end{figure}
\section{Numerical results and discussions} \label{numer}
We use the following input parameters for the numerical calculations
\cite{pdg14,yang}: \be
f_B&=&190 MeV,  f_{K_{1A}}=0.250\pm0.013 GeV, f_{K_{1B}}=0.190\pm0.01 GeV,\\
M_B&=&5.28 GeV, M_{K_{1A}}=1.31\pm0.06  GeV, M_{K_{1B}}=1.34\pm0.08 GeV\\
M_W&=&80.41 GeV, \tau_B^\pm=1.638\times 10^{-12}
s,\tau_{B^0}=1.519\times 10^{-12} s. \en For the CKM matrix
elements, we adopt the Wolfenstein parametrization and the updated
values $A=0.814, \lambda=0.22537, \bar\rho=0.117\pm0.021$ and
$\bar\eta=0.353\pm0.013$ \cite{hfag}. With the total amplitudes, one
can write the decay width as: \be \Gamma(B^+\to J/\Psi
K_1(1270,1400)^+)=\frac{G^2_F |\mathbf{P}_c|}{16\pi
m^2_B}\sum_{\sigma=L,\parallel,\perp}\mathcal{A}^\dagger_{\sigma}\mathcal{A}_{\sigma},
\en where $\mathbf{P}_c$ is the three momentum of either of the two
final state mesons, and the three helicity amplitudes are defined
as: \be \mathcal{A}_L=\mathcal{M}_L,\;\;\;\;\;
\mathcal{A}_\parallel=\sqrt2\mathcal{M}_N,
\;\;\;\;\;\mathcal{A}_\perp=r_{K_1}r_{J/\Psi}\sqrt{2(\kappa^2-1)}\mathcal{M}_T,
\en for the longitudinal, parallel, and perpendicular polarizations,
respectively, and the ratio $\kappa=P_{J/\Psi}\cdot
P_{K_1}/(M_{M_{J/\Psi}}M_{K_1})$. Then the polarization fractions
$f_\sigma(\sigma=L,\parallel,\perp)$ are written as: \be
f_\sigma=\frac{|\mathcal{A}_\sigma|^2}{|\mathcal{A}_L|^2+|\mathcal{A}_\parallel|^2+|\mathcal{A}_\perp|^2}.
\label{pol} \en With the above transversity amplitudes, one can
defined the relative phases $\phi_{\parallel}$ and $\phi_{\perp}$
as: \be
\phi_{\parallel}=arg\frac{\mathcal{A}_\parallel}{\mathcal{A}_L},
\;\;\; \phi_{\perp}=arg\frac{\mathcal{A}_\perp}{\mathcal{A}_L}.
\label{phase} \en For the charged B meson decays, the direct CP
violation $A^{dir}_{CP}$ is written as: \be
A^{dir}_{CP}=\frac{|\mathcal{\bar
A}_f|^2-|\mathcal{A}_f|^2}{|\mathcal{\bar
A}_f|^2+|\mathcal{A}_f|^2}, \label{dir} \en where $\mathcal{A}_f$ is
the total decay amplitude. If replacing $\mathcal{A}_f$ with the
different polarization amplitudes $\mathcal{A}_L,
\mathcal{A}_\parallel$ and $\mathcal{A}_\perp$, one can obtain
different direct CP violations from the different polarization
components, which are defined as $A^{dir,L}_{CP},
A^{dir,\parallel}_{CP}$ and $A^{dir,\perp}_{CP}$, respectively.

\begin{figure}[t]
\begin{center}
\vspace{-0.5cm} \centerline{\hspace{0.3cm}\epsfxsize=10.5cm
\epsffile{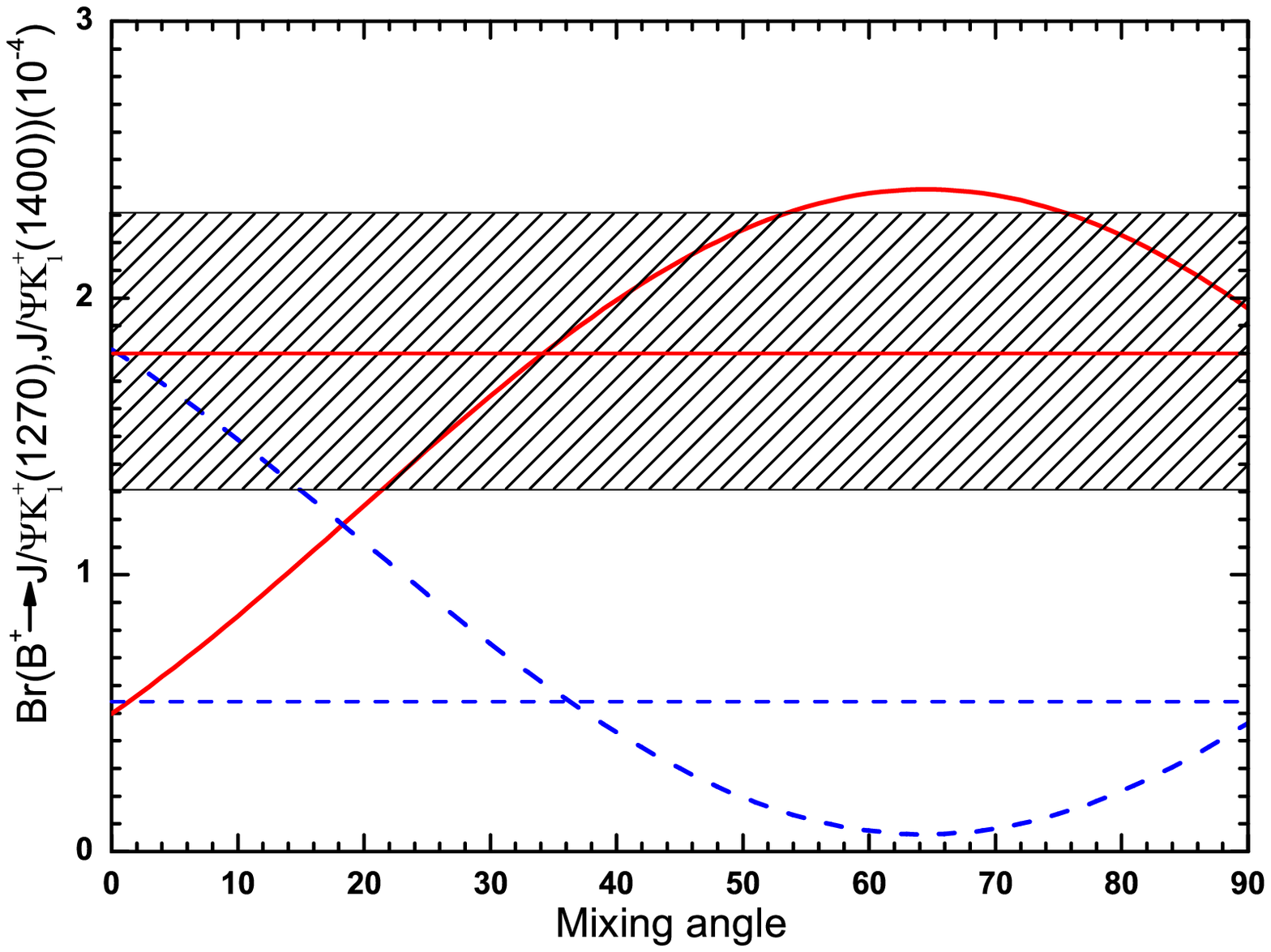} \hspace{-2.5cm} \epsfxsize=10.5cm
\epsffile{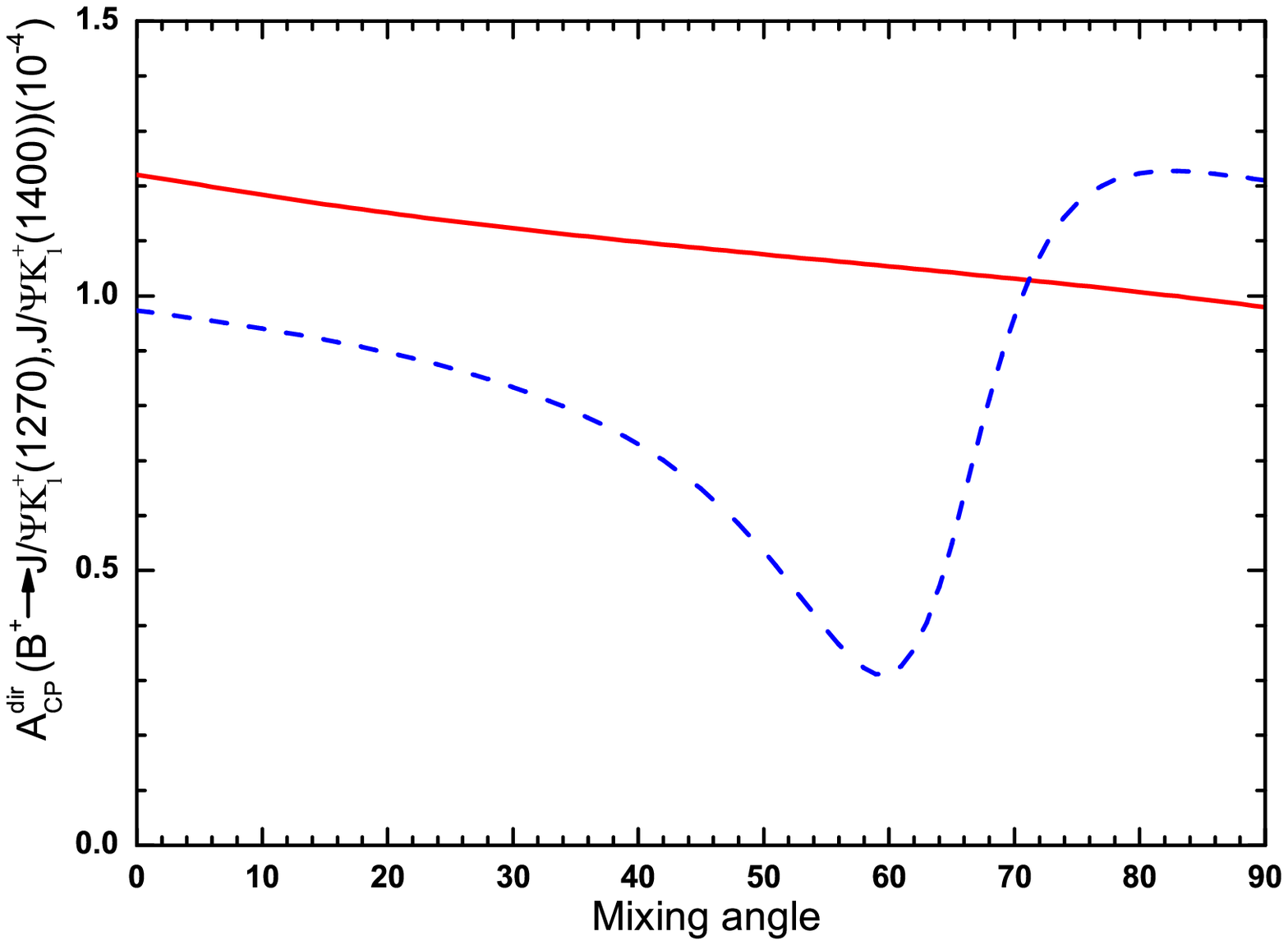}  } \vspace{-0.8cm} \caption{The (blue) dashed
curves correspond to the dependences of the branching ratio (left
panel) and the direct CP violation (right panel) on the mixing angle
for the decay $B^+\to J/\Psi K_1(1400)^+$, the (red) solid curves
refer to the dependences of the branching ratio (left panel) and the
direct CP violation (right panel) on the mixing angle for the decay
$B^+\to J/\Psi K_1(1270)^+$. On the left panel, the shaded band
shows the allowed region from the experiment and the (red)
horizontal bisector is for the central experimental value
$(1.8\pm0.5)\times10^{-3}$ of $Br(B^+\to J/\Psi K_1(1270)^+)$. The
(blue) dashed line is the upper limit for the branching ratio of
decay $B^+\to J/\Psi K_1(1400)^+$, $5.4\times10^{-4}$.}
%\vspace{-0.5cm}
\label{fig3}
\end{center}
\end{figure}
We can obtain the values of the branching ratios for decays $B^+\to
J/\Psi K_1(1270)^+$ and $B^+\to J/\Psi K_1(1400)^+$ by combining the
contributions from the flavor states $J/\Psi K_{1A}$ and $J/\Psi
K_{1B}$ through Eq.(\ref{mixing}): \be
Br(B^+\to J/\Psi K_1(1270)^+)&=&\begin{cases}(1.76^{+0.42+0.14+0.47+0.02}_{-0.57-0.13-0.36-0.02})\times10^{-3},\;\;\;\;\; \text{for}\;\;\;\;\; \theta_{K_1}=33^\circ,\\
                                           (2.36^{+0.73+0.20+0.38+0.00}_{-0.54-0.20-0.32-0.01})\times10^{-3}, \;\;\;\;\;\text{for} \;\;\;\;\;\theta_{K_1}=58^\circ,
\end{cases}
\label{bran1}
\\
Br(B^+\to J/\Psi K_1(1400)^+)&=&\begin{cases}(6.47^{+1.73+1.01+1.49+0.15}_{-1.35-0.94-1.66-0.15})\times10^{-4},\;\;\;\;\; \text{for}\;\;\;\;\; \theta_{K_1}=33^\circ,\\
                                           (8.91^{+2.85+1.77+3.56+0.12}_{-2.19-1.42-3.31-0.06})\times10^{-5}, \;\;\;\;\;\text{for} \;\;\;\;\;\theta_{K_1}=58^\circ,
\end{cases}
\label{bran2} \en where the first error comes from
$\omega_b=0.4\pm0.1$ GeV for $B$ meson, the second error is from the
decay constants $f_{K_{1A}}=0.250\pm0.013$ GeV and
$f_{K_{1B}}=0.190\pm0.01$ GeV, the third error comes from the
Gegenbauer momentums given in Appendix A, and the last one comes
from the $c$ quark mass $1.275\pm0.025$ GeV.

When the mixing angle is taken as $\theta_{K_1}=33^\circ$ , the pQCD
prediction for the decay $B^+\to J/\Psi K_1(1270)^+$ can agree well
with the experimental measurement $(1.80\pm0.52)\times10^{-3}$, at
the same time, the result for the decay $B^+\to J/\Psi K_1(1400)^+$
is near the experimental upper limit $5.4\times10^{-4}$. So we
suggest our experimental colleagues to measure carefully the
branching ratio of the decay $B^+\to J/\Psi K_1(1400)^+$ at LHCb. It
is helpful to determine the mixing angle $\theta_{K_1}$ between
$K_{1A}$ and $K_{1B}$ accurately. Considering that the difference of
the branching ratios for the neutral and charged decay modes is
mainly from the B meson lifes $\tau_{B^+}$ and $\tau_{B^0}$, one can
obtain easily the branching ratios $Br(B^0\to J/\Psi
K_1(1270)^0)=(1.63^{+0.60}_{-0.64})\times10^{-3}$ and $Br(B^0\to
J/\Psi K_1(1400)^0)=(6.52^{+2.50}_{-2.34})\times10^{-4}$ for the
mixing angle $\theta_{K_1}=33^\circ$. The former is consistent with
the experimental value $(1.30\pm0.46)\times10^{-3}$ within errors,
and the latter can be tested at the present LHCb experiment. So
comparing our predictions and the present data, one can find that
the mixing angle $\theta_{K_1}=33^\circ$ is much more favored than
$58^\circ$. In Fig.3(a), we give the dependences of the branching
ratios $Br(B^+\to J/\Psi K_1(1270)^+)$ and $Br(B^+\to J/\Psi
K_1(1400)^+)$ on the mixing angle $\theta_{K_1}$. The predictions
for the branching ratios of the decays $B^+\to J/\Psi K_1(1270)^+$
and $B^+\to J/\Psi K_1(1400)^+$ near the mixing angle $33^\circ$ can
explain the data at the same time.

\begin{table}
\caption{Our LO and NLO predictions for each polarization amplitude
(which is expressed as Pol. Amp.) for the decays $B^+\to J/\Psi
K_{1A}^+$ and $B^+\to J/\Psi K_{1B}^+$, where only the central
values are listed. The results in the brackets are the LO values,
the other results are the NLO values.   }
\begin{center}
\begin{tabular}{c|c|c|c}
\hline\hline
Decay Mode&Pol. Amp.&Tree Operators& Penguin Operators($\times10^{-2}$)\\
\hline
$B^+\to J/\Psi K_{1A}^+$&$\mathcal{M}_L$&$0.99+i0.78(1.51+i0.16)$&1.74+i1.54(8.29+i0.58)\\
\hline
$B^+\to J/\Psi K_{1A}^+$&$\mathcal{M}_N$&$0.40+i0.73(0.78+i0.22)$&$0.46+i1.57(5.51+i0.62)$\\
\hline
$B^+\to J/\Psi K_{1A}^+$&$\mathcal{M}_T$&$0.90+i1.64(1.85+0.66)$&$1.07+i3.43(12.70+i1.92)$\\
\hline\hline
$B^+\to J/\Psi K_{1B}^+$&$\mathcal{M}_L$&$0.72+i0.16(1.27-i0.66)$&$1.34+i0.41(7.72-i1.87)$\\
\hline
$B^+\to J/\Psi K_{1B}^+$&$\mathcal{M}_N$&$0.11+i0.35(0.47+i0.01)$&$0.00+i0.78(3.72+i3.56)$\\
\hline
$B^+\to J/\Psi K_{1B}^+$&$\mathcal{M}_T$&$0.03+i0.66(0.81+i0.12)$&$-0.39+i1.45(6.37+i0.21)$\\
\hline\hline
\end{tabular}\label{tab1}
\end{center}
\end{table}

When comparing the LO and NLO results, one can find that the NLO
corrections are necessary. The LO branching ratio for the decay
$B^+\to J/\Psi K_1(1270)^+$ is about $3.42\times10^{-3}$, which is
almost two times of the experimental value. After including the NLO
contributions, one can find that all of the real parts of the
amplitudes decrease consistently (shown in Table \ref{tab1}).
Furthermore, this downward trend is dominant by comparing with the
changes of each imaginary part. So the NLO branching ratio for the
decay $B^+\to J/\Psi K_1(1270)^+$ will decrease significantly and
converge with the experimental value. While the branching ratio of
the decay $B^+\to J/\Psi K_1(1400)^+$ has a tiny increase compared
with the LO result $6.38\times10^{-4}$ with the mixing angle
$\theta_{K_1}=33^\circ$.

Certainly, the mixing angle $\theta_{K_1}$ has also been checked in
other B meson decays. For example, the charged decays $B^+\to \phi
K_1(1270)^+$ and $B^+\to \phi K_1(1400)^+$ have been measured by
BaBar Collaboration \cite{babar3} with the branching ratio
$(6.1\pm1.9)\times10^{-6}$ and an upper limit $3.2\times10^{-6}$,
respectively. In order to explaining these data, many works support
the smaller mixing angle ($\sim33^\circ$) although suffering severe
interference from the annihilation type contributions. The authors
of Refs.\cite{liux,cheng1} found that the theoretical predictions
for the decay $B^+\to \phi K_1(1270)^+$ could explain the data by
taking $\theta_{K_1}\sim33^\circ$, while the values of $Br(B^+\to
\phi K_1(1400)^+)$  arrived at $10^{-5}$ order and would overshoot
the upper limit greatly. In Ref.\cite{chen1} the authors studied
these two charged decays within the generalized factorization
approach (GFA). With the annihilation type contributions turned off,
their predictions about these two channels could agree with the data
with $N^{eff}_c=5$ being the effective color number containing the
nonfactorizable effects. The similar situation also happened in the
decays $B^+\to K_1(1270)^+\gamma$ and $B^+\to K_1(1400)^+\gamma$. In
Ref.\cite{yang1} the authors explained well the data $Br(B^+\to
K_1(1270)^+\gamma)=(4.3\pm1.3)\times10^{-5}$ and $Br(B^+\to
K_1(1400)^+\gamma)<1.5\times10^{-5}$ with
$\theta_{K_1}=(34\pm13)^\circ$. Among of these decays $B\to
K_1(1270,1400)V$ ($V$ refers to a vector meson or a photon), the
branching ratios of decays $B\to K_1(1270)V$ are always larger than
those of decays $B\to K_1(1400)V$, because of the constructive
(destructive) interference between the modes $B\to K_{1A}V$ and
$B\to K_{1B}V$ through Eq.(\ref{mixing}) for the former (latter).

\begin{table}
\caption{The NLO predictions for the polarization fractions ($f_{L},
f_{\parallel}, f_{\perp}$), the direct CP violations from the
different polarization components, and the relevant phase angles
$(\phi_\parallel,\phi_\perp)$ for the decays $B^+\to J/\Psi
K_{1}(1270)^+$ with the mixing angle $\theta_{K_1}=33^\circ$ and
$58^\circ$. The first uncertainty comes from the
$\omega_b=0.4\pm0.1$ for $B$ meson, the second and the third
uncertainties are from the decay constants $f_{K_{1A}}$ and
$f_{K_{1B}}$ and the Gaigenbuar momentums in the wave functions of
$K_{1A}$ and $K_{1B}$. The last one comes from $c$ quark mass
$1.275\pm0.025$ GeV.}
\begin{center}
\begin{tabular}{c|c|c}
\hline\hline
Decay Mode&$B^+\to J/\Psi K_{1}(1270)^+$&$B^+\to J/\Psi K_{1}(1270)^+$\\
\hline
Mixing angle $\theta_{K_1}$&$33^\circ$&$58^\circ$\\
\hline
$f_L(\%)$&$52.1^{+1.4+0.3+0.9+1.2}_{-1.7-0.3-1.6-1.4}$&$49.1^{+1.0+0.2+0.5+0.9}_{-1.2-0.2-1.0-1.0}$\\
\hline
$f_\parallel(\%)$&$37.3^{+0.8+0.1+2.5+1.0}_{-0.7-0.1-1.9-0.9}$&$38.8^{+0.5+0.1+1.8+0.7}_{-0.3-0.1-1.1-0.7}$\\
\hline
$f_\perp(\%)$&$10.6^{+0.8+0.1+1.1+0.4}_{-0.7-0.1-1.2-0.3}$&$12.1^{+0.8+0.1+0.6+0.3}_{-0.7-0.1-0.9-0.3}$\\
\hline\hline
$\mathcal{A}^{L}_{CP}(10^{-4})$&$0.45^{+0.21+0.00+0.21+0.00}_{-0.21-0.00-0.23-0.01}$&$0.42^{+0.21+0.01+0.12+0.01}_{-0.20-0.00-0.26-0.02}$\\
\hline
$\mathcal{A}^{\parallel}_{CP}(10^{-4})$&$1.84^{+0.41+0.02+0.38+0.00}_{-0.38-0.02-0.36-0.01}$&$1.69^{+0.38+0.01+0.29+0.01}_{-0.35-0.01-0.20-0.01}$\\
\hline
$\mathcal{A}^{\perp}_{CP}(10^{-4})$&$1.86^{+0.22+0.03+0.32+0.02}_{-0.20-0.03-0.29-0.02}$&$1.62^{+0.19+0.01+0.26+0.01}_{-0.18-0.01-0.15-0.01}$\\
\hline\hline
$\phi_{\parallel}(rad)$&$3.98^{+0.02+0.02+0.23+0.03}_{-0.03-0.01-0.17-0.03}$&$3.78^{+0.02+0.01+0.19+0.02}_{-0.02-0.02-0.09-0.02}$\\
$\phi_{\perp}(rad)$&$4.13^{+0.01+0.03+0.24+0.04}_{-0.02-0.03-0.19-0.04}$&$3.84^{+0.01+0.01+0.18+0.03}_{-0.01-0.01-0.09-0.03}$\\
\hline\hline
\end{tabular}\label{tab2}
\end{center}
\end{table}

We also calculate the polarization fractions
$f_{\sigma}(\sigma=L,\parallel,\perp)$, the direct CP violations
$A^{dir,L}_{CP}, A^{dir,\parallel}_{CP}, A^{dir,\perp}_{CP}$ from
the different polarization components, and the relative phases
$\phi_{\parallel,\perp}$ defined in Eqs.(\ref{pol}-\ref{dir}),
respectively. The results for the decay $B^+\to J/\Psi K_1(1270)^+$
are listed in Table \ref{tab2} and for the decay $B^+\to J/\Psi
K_1(1400)^+$ in Table \ref{tab3}. Comparing with the longitudinal
polarization fractions for the decays $B^+\to J/\Psi K_1(1270)^+$
and $B^+\to J/\Psi K_1(1400)^+$, we find that the former decreases
monotonically with the increase of the mixing angle $\theta_{K_1}$
from $0^\circ$ to $90^\circ$, while the latter decreases firstly
then increases within $\theta_{K_1}\in[0^\circ,90^\circ]$. The
direct CP violation from the longitudinal component is much smaller
than those from the two transverse components for the decay $B^+\to
J/\Psi K_1(1270)^+$. As for the dependences of the total direct CP
violations for these two charged decays on the mixing angle
$\theta_{K_1}$ are shown in Fig.3(b). The total direct CP violation
values corresponding to the mixing angle $\theta_{K_1}=33^\circ$ and
$58^\circ$ are listed as following:
\begin{table}
\caption{Same as Table \ref{tab2} but with the decay $B^+\to J/\Psi
K_{1}(1400)^+$.}
\begin{center}
\begin{tabular}{c|c|c}
\hline\hline
Decay Mode&{$B^+\to J/\Psi K_{1}(1400)^+$}&$B^+\to J/\Psi K_{1}(1400)^+$\\
\hline
Mixing angle&$33^\circ$&$58^\circ$\\
\hline
$f_L(\%)$&$41.5^{+0.0+0.3+3.8+0.2}_{-0.2-0.3-4.5-0.3}$&$51.9^{+3.3+5.7+27.1+2.8}_{-3.5-4.5-28.6-2.7}$\\
\hline
$f_\parallel(\%)$&$40.9^{+0.4+0.0+4.9+0.2}_{-0.3-0.1-5.2-0.2}$&$26.3^{+2.2+4.3+25.7+1.7}_{-2.0-4.9-19.6-1.7}$\\
\hline
$f_\perp(\%)$&$17.6^{+0.4+0.4+1.8+0.1}_{-0.4-0.3-1.7-0.1}$&$21.8^{+1.3+0.3+6.9+1.0}_{-1.3-0.8-7.4-1.1}$\\
\hline\hline
$\mathcal{A}^{L}_{CP}(10^{-4})$&$0.37^{+0.11+0.00+0.24+0.05}_{-0.09-0.00-0.26-0.05}$&$0.78^{+0.26+0.07+0.32+0.11}_{-0.23-0.08-1.36-0.09}$\\
\hline
$\mathcal{A}^{\parallel}_{CP}(10^{-4})$&$1.21^{+0.28+0.03+0.46+0.02}_{-0.25-0.04-0.57-0.01}$&$-0.15^{+0.05+0.29+1.74+0.02}_{-0.07-0.33-0.68-0.00}$\\
\hline
$\mathcal{A}^{\perp}_{CP}(10^{-4})$&$0.92^{+0.12+0.04+0.37+0.00}_{-0.10-0.04-0.27-0.01}$&$-0.19^{+0.04+0.16+0.97+0.01}_{-0.03-0.17-0.93-0.01}$\\
\hline\hline
$\phi_{\parallel}(rad)$&$3.21^{+0.03+0.03+0.25+0.03}_{-0.02-0.04-0.20-0.03}$&$0.26^{+0.53+1.49+4.05+0.56}_{-0.66-5.41-4.83-0.83}$\\
$\phi_{\perp}(rad)$&$3.17^{+0.02+0.04+0.23+0.03}_{-0.02-0.04-0.17-0.03}$&$-0.08^{+0.71+1.63+3.52+0.65}_{-1.09-5.27-3.31-0.92}$\\
\hline\hline
\end{tabular}\label{tab3}
\end{center}
\end{table}
\be
A^{dir}_{CP}(B^+\to J/\Psi K_1(1270)^+)&=&\begin{cases}(1.12^{+0.27+0.00+0.10+0.01}_{-0.25-0.00-0.08-0.01})\times10^{-4},  \;\;\; \text{for} \;\;\;\theta_{K_1}=33^\circ, \\
                                                  (1.06^{+0.26+0.01+0.05+0.00}_{-0.24-0.01-0.08-0.00})\times10^{-4}, \;\;\;  \text{for} \;\;\;\theta_{K_1}=58^\circ,
\end{cases}
\\
A^{dir}_{CP}(B^+\to J/\Psi K_1(1400)^+)&=&\begin{cases}(8.11^{+1.82+0.17+1.46+0.27}_{-1.57-0.19-1.81-0.26})\times10^{-5},  \;\;\; \text{for}\;\;\; \theta_{K_1}=33^\circ, \\
                                                  (3.22^{+1.15+0.42+4.72+0.74}_{-1.17-0.21-1.00-0.76})\times10^{-5}, \;\;\;  \text{for} \;\;\;\theta_{K_1}=58^\circ,
\end{cases}
\en where the errors are the same with those in Eqs.(\ref{bran1})
and (\ref{bran2}). We adopt the Wolfenstein parametrization up to
$\mathcal{O}(\lambda^4)$ in our calculations. The weak phase will
appear in the CKM matrix element
$V_{cs}=-A\lambda^2+\frac{1}{2}A\lambda^4(1-2(\rho+i\eta))$, where
these Wolfenstein parameters are given at the start of this section.
So such small CP asymmetries are in accordance with our expectation.

\begin{table}
\caption{The NLO predictions for the branching ratios, the
polarization fractions ($f_{L}, f_{\parallel}, f_{\perp}$), the
direct CP violation, and the relevant phase angles
$(\phi_\parallel,\phi_\perp)$ for the decay $B^+\to J/\Psi
K_{1}(1270)^+$ with the mixing angle $\theta_{K_1}=33^\circ$ (top)
and $58^\circ$ (bottom), where the harmonic-oscillator type wave
functions for the $J/\Psi$ meson are used.}
\begin{center}
\begin{tabular}{cccccccccc}
\hline\hline
     &$Br(10^{-3})$&$\mathcal{A}_{CP}(10^{-4})$&$f_L(\%)$&$f_\parallel(\%)$&$f_\perp(\%)$&$\phi_{\parallel}(rad)$&$\phi_{\perp}(rad)$\\
\hline
$\omega=0.5$ GeV&$1.19$&$1.64$&$46.5$&$41.2$&$12.3$&$3.76$&$3.74$\\
$\omega=0.6$ GeV&$1.25$&$1.67$&$44.6$&$42.7$&$12.7$&$3.70$&$3.68$\\
$\omega=0.7$ GeV&$1.29$&$1.69$&$43.0$&$44.0$&$13.0$&$3.65$&$3.61$\\
\hline\hline
$\omega=0.5$ GeV&$1.55$&$1.55$&$49.8$&$37.5$&$12.7$&$3.58$&$3.53$\\
$\omega=0.6$ GeV&$1.63$&$1.58$&$48.2$&$38.8$&$13.0$&$3.55$&$3.49$\\
$\omega=0.7$ GeV&$1.69$&$1.60$&$46.8$&$40.0$&$13.2$&$3.53$&$3.46$\\
\hline\hline
\end{tabular}\label{tab4}
\end{center}
\end{table}

\begin{table}
\caption{Same as Table \ref{tab4} but for the decay $B^+\to J/\Psi
K_{1}(1400)^+$.}
\begin{center}
\begin{tabular}{cccccccccc}
\hline\hline
     &$Br(10^{-4})$&$\mathcal{A}_{CP}(10^{-4})$&$f_L(\%)$&$f_\parallel(\%)$&$f_\perp(\%)$&$\phi_{\parallel}(rad)$&$\phi_{\perp}(rad)$\\
\hline
$\omega=0.5$ GeV&$4.24$&$1.33$&$62.1$&$23.7$&$14.2$&$3.03$&$2.96$\\
$\omega=0.6$ GeV&$4.36$&$1.34$&$60.6$&$25.0$&$14.5$&$3.06$&$2.98$\\
$\omega=0.7$ GeV&$4.43$&$1.36$&$59.4$&$25.9$&$14.6$&$3.09$&$2.99$\\
\hline\hline
$\omega=0.5$ GeV&$0.58$&$1.61$&$72.3$&$13.5$&$14.2$&$4.52$&$-3.81$\\
$\omega=0.6$ GeV&$0.52$&$1.62$&$65.8$&$17.4$&$16.8$&$4.89$&$-1.67$\\
$\omega=0.7$ GeV&$0.48$&$1.62$&$59.9$&$21.3$&$18.9$&$5.48$&$-0.22$\\
\hline\hline
\end{tabular}\label{tab5}
\end{center}
\end{table}

In order to check whether the results are sensitive to the wave
functions (WFs) of $J/\Psi$ meson, we also calculate them by using
the harmonic-oscillator type wave functions for the $J/\Psi$ meson,
which are listed in Appendix A. The results for the decays $B^+\to
J/\Psi K_{1}(1270)^+$ and $B^+\to J/\Psi K_{1}(1400)^+$ are given in
Table \ref{tab4} and Table \ref{tab5}, respectively. Through
comparing these two sets of results corresponding the two type WFs
of $J/\Psi$ meson, we can see that
\begin{itemize}
\item
The branching ratios will decease about $30\%$ by using the
harmonic-oscillator type wave functions of $J/\Psi$ meson except for
that of the decay $B^+\to J/\Psi K_1(1400)^+$ with mixing angle
$\theta=58^\circ$, but anyway they keep in the same order by
changing the wave functions for $J/\Psi$ meson.
\item
For the decay $B^+\to J/\Psi K_1(1400)^+$, the polarization
fractions are sensitive to the wave functions of $J/\Psi$ meson.  If
taking the mixing angle $\theta=33^\circ$, the longitudinal
component is less than the transverse components by using Type I
WFs, but it is contrary in the case of using the harmonic-oscillator
type WFs. If taking the mixing angle $\theta=58^\circ$, the
longitudinal polarization fraction is close to the sum of other two
transverse polarization fractions in Type I WFs, while the
longitudinal polarization component is more dominant than the
transverse ones in the harmonic-oscillator type WFs.
\item
In most cases, the values of these two relative strong phases are
similar to each other in each decay mode. But for the case of the
decay $B^+\to J/\Psi K_1(1400)^+$ with the mixing angle
$\theta=58^\circ$, the relative strong phases $\phi_\parallel$ and
$\phi_\perp$ are  with opposite signs. It is valuable for us to
determine the mixing angle by measuring these relative phases from
the future experiments.
\item
In most cases, the values of the direct CP asymmetries are in the
order of $10^{-4}$ by using both of these two type WFs of $J/\Psi$
meson. But still for the case of the decay $B^+\to J/\Psi
K_1(1400)^+$ with the mixing angle $\theta=58^\circ$,  there is a
smaller direct CP violation value.
\end{itemize}
\section{Summary}
We study the B meson decays $B\to J/\Psi K_1(1270,1400)$ in the pQCD
approach beyond the leading order. With the vertex corrections and
the NLO Wilson coefficients included, the branching ratios of the
considered decays are $Br(B^+\to J/\Psi
K_1(1270)^+)=1.76^{+0.65}_{-0.69}\times10^{-3}, Br(B^+\to J/\Psi
K_1(1400)^+)=6.47^{+2.50}_{-2.34}\times10^{-4}$, and $Br(B^0\to
J/\Psi K_1(1270)^0)=(1.63^{+0.60}_{-0.64})\times10^{-3}$ with the
mixing angle $\theta_{K_1}=33^\circ$. These results can agree well
with the data or the present experimental upper limit within errors.
So we support the opinion that $\theta_{K_1}\sim33^\circ$ is much
more favored than $58^{\circ}$. We suggest our experimental
colleagues to measure carefully the branching ratio of the decay
$B^+\to J/\Psi K_1(1400)^+$ at LHCb. It is important to determine
the mixing angle $\theta_{K_1}$ between $K_{1A}$ and $K_{1B}$
accurately. On the experimental side, we find that the branching
ratios of the decays $B\to  K_1(1270)V$ ($V$ refers to a vector or a
photon) are usually much larger than those of $B\to  K_1(1400)V$. It
is because of the constructive (destructive) interference between
$B\to
 K_{1A}V$ and $B\to K_{1B}V$ for the former (latter). In order to check the dependence of our predictions on the wave functions of $J/\Psi$ meson, we also give the
results by using the harmonic-oscillator type wave functions for the
$J/\Psi$ meson, and find that these two type WFs can give the
consistent results in most cases, while some values are sensitive to
the type of wave functions of the $J/\Psi$ meson.
%%%%%%%%%%%%%%%%%%%%%%%%%%%%%%%%%%%%%%%%%%%%%%%%%%%%%%%%%%%%%%%%%%%%%%%%%%%%%%%
\section*{Acknowledgment}
This work is partly supported by the National Natural Science
Foundation of China under Grant No. 11347030, by the Program of
Science and Technology Innovation Talents in Universities of Henan
Province 14HASTIT037.
\appendix
\section{Wave functions}
For the B meson wave function, the popular parameterizations are
written as \cite{kur}: \be
\phi_B(x,b)=N_Bx^2(1-x)^2\exp\left[-\frac{m^2_Bx^2}{2\omega^2_b}-\frac{(\omega_bb)^2}{2}\right],
\en where the free paramter $\omega_b=0.40\pm0.04$ GeV and the
normalization factor $N_B=91.783$ corresponds to $\omega_b=0.40$
GeV.

For the $J/\Psi$ meson, the wave functions are given as: \be
\Psi^L(x)=\frac{1}{\sqrt{2N_c}}\left[m_{J/\Psi}\esl_L\psi^L(x)+\esl_L \psl\psi^t(x) \right],\\
\Psi^T(x)=\frac{1}{\sqrt{2N_c}}\left[m_{J/\Psi}\esl_T\psi^v(x)+\esl_T
\psl\psi^T(x) \right], \en where both the twist-2 $\psi^L(x)$ and
the twist-3 $\psi^t(x)$ will give the contribution and are listed as
\cite{bondar}: \be
\psi^L(x)&=&\psi^T(x)=9.58\frac{f_{J/\Psi}}{2\sqrt{2N_c}}x(1-x)\left[\frac{x(1-x)}{1-2.8x(1-x)}\right]^{0.7}, \label{psil}\\
\psi^t(x)&=&10.94\frac{f_{J/\Psi}}{2\sqrt{2N_c}}(1-2x)^2\left[\frac{x(1-x)}{1-2.8x(1-x)}\right]^{0.7},\\
\psi^v(x)&=&1.67\frac{f_{J/\Psi}}{2\sqrt{2N_c}}(1+(2x-1)^2)\left[\frac{x(1-x)}{1-2.8x(1-x)}\right]^{0.7}.\label{psiv}
\en where $x$ refers to the momentum fraction of the charm quark in
the charmonium meson. We call the wave functions given in
(\ref{psil}-\ref{psiv}) as Type I. Sometimes, the
harmonic-oscillator type wave functions are often used \cite{sun}:
\be
\psi^{L,T}(x,b)&=&\frac{f_{J/\Psi}N^{L,T}}{2\sqrt{2N_c}}x(1-x)\exp\left\{-\frac{m_c}{\omega}x(1-x)\left[\left(\frac{1-2x}{2x(1-x)}\right)^2+\omega^2b^2\right]\right\},\;\;\;\;\;\;\;\;\\
\psi^t(x,b)&=&\frac{f_{J/\Psi}N^t}{2\sqrt{2N_c}}(1-2x)^2\exp\left\{-\frac{m_c}{\omega}x(1-x)\left[\left(\frac{1-2x}{2x(1-x)}\right)^2+\omega^2b^2\right]\right\},\\
\psi^v(x)&=&\frac{f_{J/\Psi}N^v}{2\sqrt{2N_c}}(1+(2x-1)^2)\exp\left\{-\frac{m_c}{\omega}x(1-x)\left[\left(\frac{1-2x}{2x(1-x)}\right)^2+\omega^2b^2\right]\right\},\;\;\;\;\;\;
\en where $N^{L,T,t}$ and $N^v$ are the normalization constants and
$b$ is the conjugate variable of the transverse momentum,
$\omega=0.6\pm0.1$ GeV.

For the wave functions of the axial-vector meson $K_{1A}$ or
$K_{1B}$, they are listed as \cite{yang}: \be
\Phi^L_{K_1,\alpha\beta}&=&\langle K_1(P, \epsilon^*_L)|\bar
q_{2\beta}(z)q_{1\alpha}(0)|0\rangle \non
&=&\frac{i\gamma_5}{\sqrt{2N_c}}\int^1_0dx \; e^{ixp\cdot
z}[m_{K_1}\esl^*_L\phi_{K_1}(x)+\esl^*_L \psl\phi_{K_1}^{t}(x)
+m_{K_1}\phi^{s}_{K_1}(x)]_{\alpha\beta}, \en \be
\Phi^T_{K_1,\alpha\beta}&=&\langle K_1(P, \epsilon^*_T)|\bar
q_{2\beta}(z)q_{1\alpha}(0)|0\rangle\non
&=&\frac{i\gamma_5}{\sqrt{2N_c}}\int^1_0dx \; e^{ixp\cdot
z}\left[m_{K_1}\esl^*_T\phi^v_{K_1}(x)+\esl^*_T
\psl\phi_{K_1}^{T}(x) \right.\non && \left.
\;\;\;\;\;\;\;\;\;\;\;\;\;\;\;\;\;\;\;\;\;\;\;\;\;\;\;\;\;\;\;\;\;\;\;\;\;\;\;\;\;\;\;\;\;\;\;\;+m_{K_1}i\epsilon_{\mu\nu\rho\sigma}\gamma_5\gamma^\mu\epsilon^{*v}_Tn^\rho
v^\sigma\phi^{a}_{K_1}(x)\right]_{\alpha\beta}. \en where $K_1$
refers to the flavor state $K_{1A}$ or $K_{1B}$, and the
corresponding distribution functions can be calculated by using
light-cone QCD sum rule and listed as following: \be
\begin{cases}
\phi_{K_1}(x)=\frac{f_{K_1}}{2\sqrt{2N_c}}6x(1-x)\left[a^{\parallel}_0+3a^{\parallel}_1t+\frac{3}{2}a^{\parallel}_2(5t^2-1)\right], \\
\phi^t_{K_1}(x)=\frac{3f_{K_1}}{4\sqrt{2N_c}}\left[2a^\perp_0t^2+a^\perp_1t(3t^2-1)\right], \\
\phi^s_{K_1}(x)=\frac{f_{K_1}}{4\sqrt{2N_c}}\left[2a^\perp_1x(1-x)-a^\perp_0t-a^\perp_1t^2\right].
\label{vamp}
\end{cases}
\en The upper formulas are for the longitudinal polarization wave
functions, and the transverse polarization ones are given as: \be
\begin{cases}
\phi^T_{K_1}(x)=\frac{f_{K_1}}{2\sqrt{2N_c}}6x(1-x)\left[a^{\perp}_0+3a^{\parallel}_1t+\frac{3}{2}a^{\perp}_2(5t^2-1)\right], \\
\phi^v_{K_1}(x)=\frac{3f_{K_1}}{8\sqrt{2N_c}}\left[a^\parallel_0(t^2+1)+2a^\parallel_1t^3\right], \\
\phi^a_{K_1}(x)=\frac{3f_{K_1}}{4\sqrt{2N_c}}\left[2a^\parallel_1x(1-x)-a^\parallel_0t-a^\parallel_1t^2\right],
\label{vamp}
\end{cases}
\en where the Gegenbauer moments are given as \cite{yang, zhang1}:
\be
a^\parallel_0&=&1(-0.19\pm0.07), a^\parallel_1=-0.30^{+0.00}_{-0.20}(-1.95\pm0.45), a^\parallel_2=-0.05\pm0.03(0.10^{+0.15}_{-0.19}),\;\;\;\;\;\;\;\;\\
a^{\perp}_0&=&0.27^{+0.03}_{-0.17}(1),
a^{\perp}_1=-1.08\pm0.48(0.30^{+0.00}_{-0.33}),
a^{\perp}_2=0.02\pm0.21(-0.02\pm0.22). \en
\section{Hard functions, Evolution factors and jet functions}
The hard functions are the Fourier transformations from the
propagators of the virtual quarks and gluons, which are listed as:
\be
h_e(x_1,x_3,b_1,b_3)&=&K_0(\sqrt{x_1x_3(1-r^2_2)}m_Bb_1)\non &&\left[\theta(b_1-b_3)K_0(\sqrt{x_3(1-r^2_2)}m_Bb_1)
I_0(\sqrt{x_3(1-r^2_2)}m_Bb_3)\right.
\non && \left.+\theta(b_3-b_1)K_0(\sqrt{x_3(1-r^2_2)}m_Bb_3)I_0(\sqrt{x_3(1-r^2_2)}m_Bb_1)\right],\\
h_{d}(x_1,x_2,x_3,b_1,b_2)&=&\left[\theta(b_1-b_2)K_0(\sqrt{x_1x_3(1-r^2_2)}m_Bb_1)
I_0(\sqrt{x_1x_3(1-r^2_2)}m_Bb_2)\right.\non &&
\left.+(b_1\leftrightarrow b_2)\right]
\left(\begin{matrix}K_0(A_dm_Bb_2)& \text{for} A^2_d\geq 0\\
\frac{i\pi}{2}H^{(1)}_0(\sqrt{|A^2_d|}m_Bb_2)&\text{for} A^2_d\leq
0\\\end{matrix}\right), \en with the variables $A^2_d$ being
$A^2_d=r_c^2+(x_1-x_2)\left[(x_2-x_3)r^2_2+x_3\right]$. Here the formula for the propagator of the virtual gluons is given as $\frac{-1}{m_B^2x_1x_3(1-r^2_2)+(k_{3T}-k_{1T})^2}$.

The evolution factors are given by: \be
E_e(t)&=&\alpha_s(t)\exp[-S_B(t)-S_{K_1}(t)],\label{suda1}\\
E_{en}(t)&=&\alpha_s(t)\exp[-S_B(t)-S_{J/\Psi}(t)-S_{K_1}(t)|_{b_1=b_3}],\label{suda2}
\en where the hard scales ($t$) are chosen as: \be
t_a&=&\max(\sqrt{x_3(1-r^2_2)}m_B,1/b_1,1/b_3),\\
t_b&=&\max(\sqrt{x_1(1-r^2_2)}m_B,1/b_1,1/b_3),\\
t_{d}&=&\max(\sqrt{x_1x_3(1-r_2^2)}m_B,\sqrt{|A^2_{d}|}m_B,1/b_1,1/b_2).
\en The Sudakov exponents are defined as: \be
S_B(t)&=&s(x_1\frac{m_B}{\sqrt2},b_1)+\frac{5}{3}\int^t_{1/b_1}\frac{d\bar\mu}{\bar\mu}\gamma_q(\alpha_s(\bar\mu)),\\
S_{J/\Psi}(t)&=&s(x_2\frac{m_B}{\sqrt2},b_2)+s((1-x_2)\frac{m_B}{\sqrt2},b_2)+2\int^t_{1/b_2}\frac{d\bar\mu}{\bar\mu}\gamma_q(\alpha_s(\bar\mu)),\\
S_{K_1}(t)&=&s(x_3\frac{m_B}{\sqrt2},b_3)+s((1-x_3)\frac{m_B}{\sqrt2},b_3)+2\int^t_{1/b_3}\frac{d\bar\mu}{\bar\mu}\gamma_q(\alpha_s(\bar\mu)),
\en where the quark anomalous dimension is $\gamma_q=-\alpha_s/\pi$,
and the expression of the $s(Q,b)$ in one-loop running coupling
coupling constant is listed as: \be
s(Q,b)&=&\frac{A^{(1)}}{2\beta_1}\hat{q}\ln(\frac{\hat{q}}{\hat{b}})-\frac{A^{(1)}}{2\beta_1}(\hat{q}-\hat{b})
+\frac{A^{(2)}}{4\beta^2_1}(\frac{\hat{q}}{\hat{b}}-1)\non
&&-\left[\frac{A^{(2)}}{4\beta^2_1}-\frac{A^{(1)}}{4\beta_1}\ln(\frac{e^{2\gamma_E-1}}{2})\right]
\ln(\frac{\hat{q}}{\hat{b}}), \en here the variables are defined by
$\hat{q}=\ln[Q/(\sqrt2\Lambda)], \hat{b}=\ln[1/(b\Lambda)]$ and the
coefficients $A^{(1,2)}$ and $\beta_{1}$ are given as:\be
\beta_1&=&\frac{33-2n_f}{12},A^{(1)}=\frac{4}{3},\\
A^{(2)}&=&\frac{67}{9}-\frac{\pi^2}{3}
-\frac{10}{27}n_f+\frac{8}{3}\beta_1\ln(\frac{1}{2}e^{\gamma_E}),
\en where $n_f$ is the number of the quark flavors and $\gamma_E$
the Euler constant.
\section{Vertex functions}
The hard scattering functions $f_I$ and $g_I$ arised from the vertex
corrections are given as \cite{cheng11,liu}: \be
f_I&=&\frac{2\sqrt{2N_C}}{f_{J/\Psi}}\left\{\int^1_0
dx_2\psi^{L}_{J/\Psi}(x_2)\left[\frac{2r^2_2x_2}{1-r^2_2(1-x_2)}
+(3-2x_2)\frac{\ln x_2}{1-x_2}\right.\right.\non
&&\left.\left.+\left(-\frac{3}{1-r^2_2x_2}+\frac{1}{1-r^2_2(1-x_2)}-\frac{2r^2_2x_2}{[1-r^2_2(1-x_2)]^2}\right)
r^2_2x_2\ln(r^2_2x_2)\right.\right.\non
&&\left.\left.+\left(3(1-r^2_2)+2r^2_2x_2+\frac{2r^4_2x^2_2}{1-r^2_2(1-x_2)}\right)
\frac{\ln(1-r^2_2)-i\pi}{1-r^2_2(1-x_2)}\right]\right.\non &&\left.
+\int^1_0dx_2\psi^{T}_{J/\Psi}(x_2)\left[\frac{-8x^2\ln
x}{1-x}+\frac{8r^2_2x^2\ln(r^2_2x)}{1-r^2_2(1-x)}-8r^2_2x^2\frac{\ln(1-r^2_2)-i\pi}
{1-r^2_2(1-x)}\right]\right\},\\
g_I&=&\frac{2\sqrt{2N_c}}{f_{J/\Psi}}\left\{\int^1_0dx\psi^{L}_{J/\Psi}(x)\left[\frac{-4x\ln
x}{(1-r^2_2)(1-x)}+
\frac{r^2_2x\ln(1-r^2_2)}{[1-r^2_2(1-x)]^2}+r^2_2x\ln(r^2_2x)\right.\right.\non
&&\left.\left.
\times\left(\frac{1}{(1-r^2_2x)^2}-\frac{1}{[1-r^2_2(1-x)]^2}+\frac{2(1+r^2_2-2r^2_2x)}{(1-r^2_2)(1-r^2_2x)^2}\right)
-\frac{i\pi r^2_2x}{[1-r^2_2(1-x)]^2}\right]\right.\non &&\left.
\int^1_0\psi^{T}_{J/\Psi}(x)\left[\frac{8x^2\ln
x}{(1-r^2_2)(1-x)}-\frac{8x^2r^2_2\ln(r^2_2x)}{(1-r^2_2)(1-r^2_2x)}\right]\right\}.
\en
%%%%%%%%%%%%%%%%%%%%%%%%%%%%%%%%%%%%%%%%%%%%%%%%%%%%%%%%%%%%%%%%%%%%%%%%
%                               references
%%%%%%%%%%%%%%%%%%%%%%%%%%%%%%%%%%%%%%%%%%%%%%%%%%%%%%%%%%%%%%%%%%%%%%%%

\end{document}